\documentclass[sigconf]{acmart}

\setcopyright{none}  
\renewcommand\footnotetextcopyrightpermission[1]{}  

\acmConference[Augmented Educators and AI(CHI 2025 Workshop)]{Augmented Educators and AI: Shaping the Future of Human-AI Collaboration in Learning on CHI 2025 Workshop}{April 26,2025}{Yokohama, JAPAN}

\begin{document}

\title{Educational Twin: The Influence of Artificial XR Expert Duplicates on Future Learning}

\author{Clara Sayffaerth}
\affiliation{%
  \institution{LMU Munich}
  \city{Munich}
  \country{Germany}}
\email{clara.sayffaerth@ifi.lmu.de}

\renewcommand{\shortauthors}{Sayffaerth}

\begin{abstract}
  Currently, it is impossible for educators to be in multiple places simultaneously and teach each student individually. Technologies such as Extended Reality (XR) and Artificial Intelligence (AI) enable the creation of realistic educational copies of experts that preserve not only visual and mental characteristics but also social aspects crucial for learning. However, research in this area is limited, which opens new questions for future work. This paper discusses how these human digital twins can potentially improve aspects like scalability, engagement, and preservation of social learning factors. While this technology offers benefits, it also introduces challenges related to educator autonomy, social interaction shifts, and ethical considerations such as privacy, bias, and identity preservation. We outline key research questions that need to be addressed to ensure that human digital twins enhance the social aspects of education instead of harming them.
\end{abstract}


\begin{teaserfigure}
  \includegraphics[width=\textwidth]{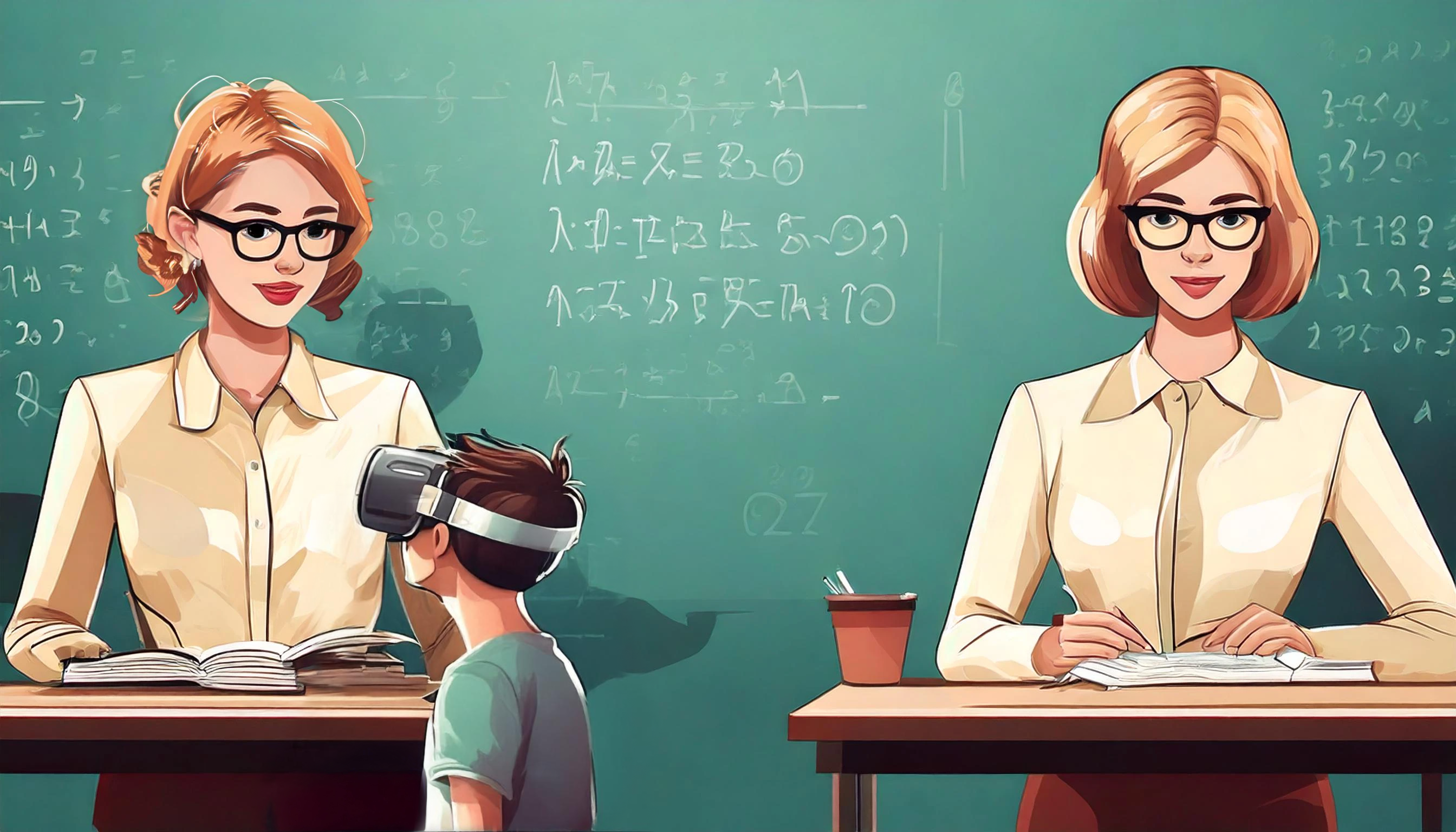}
  \caption{An educational twin is giving a student individual tutoring in XR while the original expert is preparing the next lecture (\textit{created with Firefly})}
  \label{fig:teaser}
\end{teaserfigure}

\maketitle

{\small
\noindent ©  This paper was adapted for the \textit{CHI 2025 Workshop on Augmented Educators and AI: Shaping the Future of Human and AI Cooperation in Learning},
held in Yokohama, Japan on April 26, 2025. This work is licensed under the Creative Commons Attribution 4.0 International License (CC BY 4.0).
}

\section{Introduction}
Teachers usually educate large groups of students and rarely have time for individual support due to staff shortages and the resulting workload~\cite{malmComplexitiesEducatingStudent2020, otsukiAssessmentInstructorCapacity2022}. While some students can cope with this situation or receive expensive external tutoring, others need alternative additional support to avoid failure. 

To overcome these problems in large-scale learning environments, various technological aids are integrated into teaching, such as video recordings of lectures or educational computer games to review the learned material. Although these approaches are valuable, challenges remain in maintaining the important social aspects of teaching~\cite{banduraSociallearningTheoryIdentificatory1969} such as trust and empathy. These are difficult to replicate and build, as they often develop over time through direct interactions between the teacher and the student. For example, practical and procedural knowledge is traditionally taught by educators through demonstration and hands-on learning~\cite{brunerFolkPedagogies1999, bretzSelectionAppropriateCommunication1971}, making it difficult to explain them using linear two-dimensional methods such as video recordings. In these situations, the loss of social and emotional elements can potentially hinder the learning experience~\cite{rodriguez-keyesBeingKnownUndergraduate2013}.

In research and industry, digital twins are often used to preserve the properties, states, and individual characteristics of objects and to transfer a copy of them into the virtual space. These duplicates can be useful for data exchange and simulations. In the educational context, this approach could allow teachers to essentially split themselves and provide individualized help, anywhere and anytime. Such educational twins go beyond simple recordings and playback of digital content, but are independent instances that generate individual and personalized learning content while looking, behaving, and interacting with students like the original. Technologies like Extended Reality (XR)~\cite{caoExploratoryStudyAugmented2020} and Artificial Intelligence (AI)~\cite{ifelebueguChatbotsAIEducation2023} can not only visualize the educators outside but also, through trained models, preserve their inner mental characteristics and use them independently of time and place. This potentially creates social connections in the absence of physical presence, which can be crucial in the learning context. 

While such developments have advantages, these educational twins also pose challenges. This paper examines the potential of AI and XR to create human digital twins, explores related research, and highlights the open questions that should be answered to utilize the benefits of these technologies while preventing their risks. Through this exploration, we aim to provide insights into how AI and XR can enhance the future of human-AI collaboration in education.

\section{Related Work}
Teaching a large group of students poses challenges not only in real educational institutions~\cite{biggsWhatStudentDoes1999} but also in online courses~\cite{zhengUnderstandingStudentMotivation2015}. Collaboration between several students at the same time can therefore increase the teacher's workload~\cite{otsukiAssessmentInstructorCapacity2022}, potentially leading to negative effects on the quality of teaching. Without additional help, this can be particularly harmful for already struggling students~\cite{francisTeacherQualityandAttainment2019}. For this reason, new applications are constantly being created to support learners individually.

In addition, the rapid development of large language models (LLMs) and AI has significantly changed the way we learn. AI educational tools can generate adaptive learning experiences, personalize feedback, and automate instructional support~\cite{labadzeRoleAIChatbots2023}. At the same time, technologies such as 3D scanning avatars~\cite{shenXAvatarExpressiveHuman2023, linTransformingEngineeringEducation2024}, volumetric videos~\cite{irlittiVolumetricMixedReality2023}, and human digital twins~\cite{nguyenExploratoryModelsHumanAI2024} are evolving to enable realistic representations of people in XR environments. These developments create new opportunities for immersive and interactive learning by enabling students to engage with AI-driven visualizations that resemble real educators in both appearance and behavior~\cite{vallisStudentPerceptionsAIGenerated2024}.

While research has mostly looked at AI-driven adaptive learning and XR-based educational environments separately, few studies have explored the combination of both technologies to support educators: Existing work has demonstrated how AI could enhance XR learning environments by making avatars more adaptive to user needs, for example, through guiding the learners~\cite{khokharModifyingPedagogicalAgent2022} or making virtual environments more interactive~\cite{liarokapisExtendedRealityEducational2024, gaoAIDrivenAvatarsImmersive2024}. 

However, there is a gap in assessing the social aspects of interacting with AI avatars in XR learning environments. These social interactions play a crucial role in education, as they influence motivation, engagement, and knowledge retention. Little research has been conducted on how realistic-looking AI copies of educators in XR can affect trust, empathy, and social dynamics in the learning process. Further research is needed to bridge the gap between AI personalization and socially meaningful interactions in XR. Investigating these aspects is essential to ensure that AI-XR educators not only provide effective instruction by fostering the social and emotional connections necessary for meaningful learning experiences but also cause no harm.

\section{Opportunities and Challenges}

In the following section, we will go into detail on the opportunities and challenges that need to be addressed in order to facilitate social and safe learning with these technologies in the future.

\subsection{Didactic Scalability}
The integration of AI and XR technologies in education offers advances in didactic scalability, making learning more accessible in terms of time and place. 
For example, students who need additional time to learn a particular subject can interact with the model wherever, whenever, and for as long as they wish while also being able to adapt aspects such as the spoken language~\cite{al-dargazelliHPLCTrainingAll2024}, leading to potentially equal opportunities for all students.
In addition, as already mentioned above, these technologies can increase engagement by making learning experiences more interactive and personalized. This leads to less stress for educators by outsourcing repetitive teaching tasks.

However, with these benefits come challenges. Questions of ownership and control of three-dimensional AI-generated teacher models also arise: who owns, maintains, and updates these models, and how can uncontrolled sharing and modification of the digital copy be prevented to protect the identity of the educational original? As AI tools such as LLMs are still prone to error, the flawed characteristics of an educational copy can also affect the perception of the real educator. As a result, the reputation of the expert can suffer or, if used on a large scale, even the whole profession. This phenomenon has already been observed in other areas, such as the increased gender bias through the use of female voice assistance~\cite{mahmoodGenderBiasesError2024}. Students should therefore learn to critically engage with and reflect on such tools. In addition, over-reliance on AI-driven teaching could reduce educators’ leadership skills, adaptability, and pedagogical flexibility, potentially leading to job insecurity if institutions prioritize low-cost alternatives over human educators~\cite{ghamrawiExploringImpactAI2024}. Research should also explore how to keep experts in the loop to avoid such incidents. The amount of hardware and computing power required to train and run such models can not only harm the environment, but may also not be affordable for everyone, creating additional disadvantages for certain institutions and student groups.

\subsection{Social Aspects}
A key argument in favor of AI- and XR-enhanced education is its ability to preserve the social aspects that are important for learning~\cite{khokharModifyingPedagogicalAgent2022, thanyaditTutorInsightGuiding2023}. Unlike purely text-based or asynchronous supporting methods, AI educator models can maintain a social presence through their realism and interactivity, fostering stronger emotional connections with students and potentially improving learning outcomes.

However, significant concerns emerge about the nature of these social interactions. While AI-generated educators facilitate learning, they introduce a new layer of social dynamics that differ from human relationships. For example, the social connection built with an AI educator is not experienced by the original human educator, raising questions about how real-life teacher-student interactions will change in the future. In addition, interactions with AI could reshape social behavior not only in how students interact with teachers, but also in how they interact with people in general. These changes may, over time, affect empathy, trust, and perceptions of authority. Furthermore, the behavioral complexity of educators may not be easily replicated in AI systems. People´s personal characteristics and motivations, including their adaptive decision making, non-verbal cues, and nuanced instructional adaptations, as well as their experiences and preferences, influence how they behave and react in their environment and therefore how they transfer knowledge. So personal data must also be considered in order to create a realistic representation, which could lead to privacy violations. It is also unclear which of these social aspects helps improve the learning outcomes. Another challenge is the potential bias embedded in AI educator models: If an AI system reflects the opinions or unconscious biases of a real educator, or is manipulated to do so, it could amplify inequalities and negatively affect certain student groups on a large scale. As a result, future systems need to be designed in a way that they treat all learners equally.

\subsection{Version Differences}


As the generated educator models can also be adapted compared to the original, they can be used to provide consistent instruction, ensuring that learners receive the same quality of knowledge delivery regardless of external constraints such as sympathy or bias towards different students. As a result, every student can receive the same treatment and have the opportunity to learn according to their specific needs~\cite{mclarenPoliteWebbasedIntelligent2011}. The knowledge of the digital educator can easily be preserved, expanded, and linked to large knowledge databases and the Internet, bringing additional benefits to learners.

When changing the model, the question arises as to whether an adapted version of the educator is still seen as the original, and how the copied experts feel when they are altered. A characteristic of AI-driven educational models is that they visually and mentally represent an educator at a fixed point. While this allows for the preservation of teaching styles, methods, and knowledge, AI models can continue to evolve autonomously and individually for each person, potentially diverging from the original educator’s philosophy and pedagogical intent over time. On the one hand, without regular updates, these models may no longer resemble the original educator, raising questions about authenticity and identity preservation. On the other hand, updating the models may change each student's version, resulting in a potential loss of social connection due to changes in characteristics. This can lead to frustration and reduced learning success.

Furthermore, AI-generated educators rely on extensive data collection, raising questions about ethics, data privacy, informed consent, and ownership of educational materials and interactions~\cite{liarokapisExtendedRealityEducational2024}. Additionally, the perceived reliance on AI models can influence students in unintended ways. If they are trained on the data of a single educator, they may reinforce specific viewpoints on a wide range of people. These challenges highlight the need for transparency, ethical AI governance, and clear policies to ensure that AI-driven education remains fair, effective, and aligned with human values.

\section{Conclusion}
AI and XR technologies offer potential for education by improving scalability, accessibility, and engagement. These technologies can preserve and even enhance social aspects of learning while providing personalized instruction. However, challenges remain, including concerns about social interactions, ethical considerations, privacy, and the evolving nature of AI-driven educator models.

To benefit from the combination of AI and XR education, further research is needed to understand its long-term impact on learning dynamics, social behavior, and educator-student relationships. In addition, frameworks for ethical AI governance need to be established to ensure responsible implementation. By addressing these open questions, we can create AI educational environments that complement rather than replace human educators, fostering meaningful and equal learning experiences for all students.

\section{The Author}
Clara Sayffaerth is a PhD student at LMU Munich, supervised by Albrecht Schmidt. Her research focuses on the use of XR and AI for knowledge transfer. Together with the Department of Physics Education and various museums, including her former employer, the Deutsches Museum\footnote{\protect\url{https://www.deutsches-museum.de/en}, Last Accessed: \today}, she is developing solutions to improve learning experiences while identifying potential risks.


\bibliographystyle{ACM-Reference-Format}
\bibliography{Asynchronous_Assistence}

\end{document}